\documentstyle[preprint,aps]{revtex}

\begin{document}
\draft
                     
\title{Comment on `` Cosmological Gamma Ray Bursts and the Highest Energy
Cosmic Rays''} 

\author{Arnon Dar}
\address{Theory Division, CERN, CH-1211 Geneva 23, Switzerland\\  
and\\ 
Department of Physics and
Space Research Institute\\ 
Technion, Israel Institute of Technology, Haifa 32000, Israel}

\maketitle

In a letter with the above title, published some time ago in PRL, Waxman
[1] made the interesting suggestion that cosmological
gamma ray bursts (GRBs) are the source of the ultra high energy cosmic
rays (UHECR). This has also been proposed independently by Milgrom and
Usov [2] and by Vietri [3]. However, here I show that the recent data from
AGASA [4], which does not show the suppression in the intensity of cosmic
rays (CR) at energy beyond$ \sim 4\times 10^{19}eV$ that was predicted by
Greisen, Zatsepin and Kuz'min (GZK)  for extragalactic cosmic rays due to
their interaction with the cosmic background photons [5], rules out
extragalactic GRBs as the source of UHECR.

For simplicity, consider first a uniform  distribution of galaxies
(or sites of GRBs) in Euclidean space, with a number density $n$ 
per unit volume. Let $\Delta E_{CR}$ be the mean energy release in CR  per
GRB
and $R_G$ the GRB rate per galaxy. Neglecting scattering and
attenuation in the intergalactic space, the contribution of such sources 
within a distance $D$ from Earth to the observed CR energy flux
(energy per unit area, per $sr$, per unit time) is   
\begin{equation}
S\approx {nR_G\Delta E_{CR}D\over 4\pi}~.
\end{equation}
Cosmic expansion and evolution are unimportant  for
cosmological distances  $D\ll c/H\sim
3000h^{-1}~Mpc$ where $H=100h~km~s^{-1}~Mpc^{-1}$.  If UHECR are trapped
locally by (unknown) strong extragalactic magnetic fields that surround
our Milky Way galaxy, then $D$ in eq. 1 must be replaced by $c\tau(E)$,
where
$\tau(E)$ is the lifetime of UHECR in the trap due to attenuation and/or
escape by diffusion in the magnetic fields. The mean attenuation length
(lifetime) of CR with energies above $4\times 10^{19}eV$ 
(the GZK ``cutoff'' energy) due to
photoproduction in collisions with the cosmological background photons
yields [6] $D<15~Mpc$ ( $\tau <5\times 10^8~y$).  
The attenuation 
of CR protons below the GZK ``cutoff'' energy 
is mainly due to $e^+e^-$ pair
production on the cosmic background photons. The attenuation length 
(time) Just below 
the GZK ``cutoff'' energy is [6] $D\sim 1000~Mpc$ ($\tau
\sim 3\times 10^{10}~y$). It increases  with decreasing energy to
the Hubble radius (Hubble time) at about $10^{18.5}eV$, the energy of the
CR ankle. Since $D$ $(c\tau)$ changes at the GZK ``cutoff'' energy by a
factor $\geq 1000/15\sim 70$,  the  intensity
of the UHECR flux beyond the GZK ``cutoff'' is  suppressed by  
the same factor.  Such a suppression was not observed by AGASA [4].

\noindent Moreover, GRBs that emit isotropically and are uniformly
distributed within $D<20~Mpc$ cannot produce the measured intensity of
UHECR above the CR ankle: The measured luminosity density in the local
Universe is [7] $n\sim 1.8h\times 10^8L_\odot~Mpc^{-3}$.  From the
observed rate of cosmological GRBs [8], $R_{obs}\sim 10^3~y^{-1}$, and
from the recently measured/estimated redshifts of some GRBs and host
galaxies of GRBs [9], it was concluded that [10] $R_*< 10^{-8}~y^{-1}$ per
$L_*\sim 10^{10}L_\odot$ galaxy. The kinetic energy release in UHECR per
birth/death of compact stellar objects, which are the plausible triggers
of GRBs in galaxies, probably, does not exceed [11] $\Delta E_{CR}\sim
10^{52}~erg$ (Waxman's assumption, $\Delta E_{CR}=5\epsilon \times 10^{50}
~erg$ where $\epsilon$ is the mean energy of UHECR in $10^{20}eV$ units,
makes the discrepancy even larger).  Thus, for $h\sim 0.65$, eq. 1 yields
an energy flux of UHECR, \begin{equation} S\sim 30\left ({n\over
1.8h\times 10^{-2}Mpc^{-3}}\right ) \left ({R_G\over 10^{-8}y^{-1}}\right
)\left ( {\Delta E_{CR}\over 10^{52}erg}\right )  \left ( {D\over
20Mpc}\right )~eV~m^{-2}s^{-1}sr^{-1}.  \end {equation} The CR above the
ankle have an approximate power-law spectrum [4] $dn/dE\approx
AE^{-\beta}$ with $\beta\sim 2.5$~.  Even in the very unlikely situation
where the bulk of the GRB energy is carried by CR with energy above
$E_0\sim 10^{20}eV$, one obtains from eq. 2 that for $E\sim E_0$
\begin{equation} E^3{dn\over dE}\sim (\beta-2)SE_0\left ( {E\over
E_0}\right )^{1-\beta} \sim 3\times 10^{21}eV^2m^{-2}s^{-1}sr^{-1}. 
\end{equation} The observed value [4], $E^3dn/dE\sim 4\times
10^{24}eV^2m^{-2}s^{-1}sr^{-1}$ around $E\sim E_0$, is smaller by three
orders of magnitude than that predicted by eq. 3. \noindent The above
luminosity problem can be solved by postulating that GRBs emit
isotropically more than $10^{55}$ erg in UHECR (very unlikely), or by
jetting the GRB ejecta [11]. If the GRB ejecta is collimated into a narrow
jet (plasmoid) with a bulk motion Lorentz factor $\Gamma\sim 10^3$ then
its radiation is beamed into a solid angle $\Delta\Omega\sim
\pi/\Gamma^2$. The ``isotropic'' energy release, $E_{isot}\equiv 4\pi
(\Delta E/\Delta \Omega)= (4\pi/\Delta \Omega)\Delta E \sim 4\times 10^6
(\Gamma/10^3)^2 \Delta E$, can be much larger than $\Delta E $, the true
energy release in GRB. Thus, while the total luminosity,
$R_{obs}E_{isot}$, from GRBs, is independent of the beaming angle,
$E_{isot}$ can be much larger than that assumed by Waxman [1] and Vietri
[3]. However, extragalactic GRBs cannot smear the GZK ``cutoff`` unless
there is a ``cosmic conspiracy'', namely, the large scale local magnetic
fields conspire to trap the extragalactic UHECR at the GZK ``cutoff''
energy for a time which is approximately equal to their attenuation time
in the background radiation [12].  Such a possibility also seems very
remote in view of observational limits on extragalactic magnetic fields
from limits on Faraday rotation of radio waves from distant powerful radio
sources [13] and from limits on intergalactic synchrotron emission: The
Larmor radius of $4\times 10^{19}eV$ protons in typical extragalactic
magnetic fields, $(B<10^{-9}G)$, is much larger than the typical coherence
length $(l_c<1~Mpc)$ of these fields. Moreover, magnetic trapping is
completely ruled out if the arrival directions of UHECR coincide with the
directions of cosmological GRBs [2] or if the arrival directions of
extragalactic UHECR are clustered [14]. Thus, if extragalactic CR protons
below the GZK ``cutoff'' suffer only small random magnetic deflections
through intergalactic space and reach us from distances as far as $D\sim
1000~Mpc$ away, then their flux must show the the GZK suppression when
their energy increases beyond $\sim 4\times 10^{19}eV$. 

The same arguments can be repeated for cosmic ray nuclei which are
attenuated via photodissociation by cosmic background radiations. They
lead to the same conclusion, namely, the UHECR which are observed near
Earth, most probably, are not produced by extragalactic GRBs. 
However, narrowly collimated Galactic GRBs, most of which do not point
in our direction, can be the  source  of non solar cosmic ray nuclei at
all energies [15,16].

\end{document}